\documentclass[letter,apj]{emulateapj-rtx4}
\usepackage{graphicx}
\usepackage{enumerate}
\usepackage{amssymb, amsmath}
\usepackage{mathtools}
\usepackage{natbib}
\usepackage{color}
\usepackage{ulem}
\usepackage{soul}
\usepackage{bm}
\newcommand{\ias}{2}
\newcommand{\princeton}{1}
\newcommand{\sagan}{3}

\begin{document}

\title{The measurement, treatment, and impact of spectral covariance and Bayesian priors in integral-field spectroscopy of exoplanets}
\author{Johnny P. Greco\altaffilmark{\princeton} and
Timothy D.~Brandt\altaffilmark{\ias,\sagan}
}

\altaffiltext{\princeton}{Department of Astrophysical Sciences, Princeton University, Princeton, NJ, USA}
\altaffiltext{\ias}{Institute for Advanced Study, Princeton, NJ, USA}
\altaffiltext{\sagan}{NASA Sagan Fellow}

\begin{abstract}
The recovery of an exoplanet's atmospheric parameters from its spectrum requires accurate knowledge of the spectral errors and covariances. Unfortunately, the complex image processing used in high-contrast integral-field spectrograph (IFS) observations generally produces spectral covariances that are poorly understood and often ignored. In this work, we show how to measure the spectral errors and covariances and include them self-consistently in parameter retrievals. By combining model exoplanet spectra with a realistic noise model generated from GPI early science data, we show that ignoring spectral covariance in high-contrast IFS data can both bias inferred parameters and lead to unreliable confidence regions on those parameters. This problem is made worse by the common practice of scaling the $\chi^2$ per degree of freedom to unity; the input parameters then fall outside the 95\% confidence regions in as many as ${\sim}80\%$ of noise realizations. The biases we observe can approach the typical levels of precision achieved in high-contrast spectroscopy. Accounting for realistic priors in fully Bayesian retrievals can also have a significant impact on the inferred parameters. Plausible priors on effective temperature and surface gravity can vary by an order of magnitude across the confidence regions appropriate for objects with weak age constraints; priors for objects with good age constraints are dominated by modeling uncertainties. Our methods are directly applicable to existing high-contrast IFSs including GPI and SPHERE, as well as upcoming instruments like CHARIS and, ultimately, WFIRST-AFTA. 
\end{abstract}


\section{Introduction} \label{sec:intro}

Large ground-based telescopes with adaptive optics have now imaged tens of substellar companions with masses ranging from that of massive brown dwarfs to ${\sim}3~M_{\rm Jup}$ giant planets \citep[e.g.,][]{Marois+Macintosh+Barman+etal_2008,Thalmann+Carson+Janson+etal_2009,Kuzuhara+Tamura+Kudo+etal_2013,Rameau+Chauvin+Lagrange+etal_2013,Macintosh+Graham+Barman+etal_2015, Wagner+Apai+Kasper+etal_2016}.  Direct imaging collects photons emitted by the planets and brown dwarfs, allowing them to be characterized by their colors \citep{Burrows+Marley+Hubbard+etal_1997,Knapp+Leggett+Fan+etal_2004,Saumon+Marley_2008,Skemer+Marley+Hinz+etal_2014}, and ultimately by their chemistry and composition \citep{Bowler+Liu+Dupuy+etal_2010,Barman+Macintosh+Konopacky+Marois_2011a,Janson+Brandt+Kuzuhara+etal_2013,Konopacky+Barman+Macintosh+etal_2013}. A planet's temperature, mass, and age can be used to infer its formation conditions \citep{Marley+Fortney+Hubickyj+etal_2007,Spiegel+Burrows_2012}, while its atmospheric abundances might show where in the protoplanetary disk it accreted its atmosphere \citep{Oberg+Murray-Clay+Bergin_2011,Madhusadhan_2012}. The field of high-contrast imaging has expanded dramatically in recent years and continues to push to higher contrasts and smaller separations \citep[][and references therein]{Bowler_2016}.

High-contrast imaging has long relied on observing strategies and image processing to achieve a factor of ${\sim}$10 improvement over an instrument's raw contrast.  Surveys with the {\it Hubble Space Telescope} ``rolled'' the spacecraft to rotate the point-spread function (PSF) with respect to the field, subtracting the PSF at different orientations \citep{Schneider+Silverstone_2003}.  Angular differential imaging (ADI) enables the same technique to be used in ground-based imaging \citep{Marois+Lafreniere+Doyon+etal_2006}, with additional image processing to model and subtract the stellar PSF \citep{Lafreniere+Marois+Doyon+etal_2007,Soummer+Pueyo+Larkin_2012}. This image processing partially subtracts the flux from faint companions, with the magnitude of the effect varying with position. The partial subtraction is usually estimated by introducing and reducing fake point sources \citep[e.g.][]{Lafreniere+Marois+Doyon+etal_2007}, though it may also be forward-modeled as a perturbation to the data reduction \citep{Brandt+McElwain+Turner+etal_2013,Pueyo_2016}. An alternative observing strategy known as spectral differential imaging (SDI) simultaneously images a star in two or more neighboring wavelengths.  The images are then subtracted to remove starlight; companions with strong spectral features will remain in the residual images \citep{Racine+Walker+Nadeau+etal_1999,Marois+Doyon+Racine+etal_2000}.  SDI has been used in high-contrast surveys with NICI on Gemini South \citep{Biller+Liu+Wahhaj+etal_2013,Nielsen+Liu+Wahhaj+etal_2013,Wahhaj+Liu+Biller+etal_2013} and with NaCo on the VLT \citep{Biller+Close+Maciadri+etal_2007,Maire+Boccaletti+Rameau+etal_2014}.  Unfortunately, the contrast of SDI imaging is a strong function of the companion's spectrum.

The new generation of high-contrast instruments are integral-field spectrographs \citep[IFSs; e.g.,][]{Beuzit+Feldt+Dohlen+etal_2008,Macintosh+Graham+Palmer+etal_2008,Hinkley+Oppenheimer+Zimmerman+etal_2011,Peters+Groff+Kasdin+etal_2012}, which promise better performance but pose a new set of data analysis challenges.  An IFS simultaneously obtains a spectrum of each spatial element in its field-of-view, producing a three-dimensional data cube composed of one spectral ($\lambda$) and two spatial ($x$, $y$) dimensions.  The stellar PSF can now be modeled and subtracted in three dimensions; image processing can exploit the scaling of diffraction with wavelength to improve the instrument's raw contrast beyond what is possible with ADI imaging \citep{Sparks+Ford_2002,Berton+Gratton+Feldt+etal_2006,Marois+Correia+Galicher+etal_2014}.  IFSs also naturally enable the recovery of an exoplanet's spectrum.  With current image processing techniques, the spectrum is subject to the same partial subtraction biases as high-contrast photometry.  It also suffers from spectral covariance.  The stellar PSF scales with wavelength but is coherent on spatial scales of $\lambda/D$; errors in the model PSF couple neighboring wavelengths in the exoplanet's spectrum.

Exoplanet spectra obtained with IFSs have shown clear signs of clouds and non-equilibrium chemistry in the atmosphere of HR8799~b \citep{Bowler+Liu+Dupuy+etal_2010, Barman+Macintosh+Konopacky+Marois_2011a}, as well as the presence of carbon monoxide and water in the atmosphere of HR8799~c \citep{Konopacky+Barman+Macintosh+etal_2013}.  The large grids of substellar model atmospheres now available from several groups \citep{Allard+Homeier+Freytag_2011,Spiegel+Burrows_2012} may be used in a global fit to the spectrum \citep{Rice+Oppenheimer+Zimmerman+etal_2015}.  An alternative approach is to fit the abundances and temperature profiles of the most important species to the observed spectrum \citep{Line+Teske+etal_2015}. Either approach, using grids of atmospheres that assume equilibrium chemistry and (typically) Solar abundance patterns or by fitting molecular species individually, requires the spectral errors and covariances to be properly modeled and understood.

In the high-contrast regime, it becomes challenging to understand the effects of data analysis techniques on the recovered spectra. The Gemini Planet Imager (GPI) recently discovered a low-mass companion to the young F-type star 51 Eridani \citep{Macintosh+Graham+Barman+etal_2015}. The authors of this study extracted the spectrum of 51 Eri b using three different data analysis pipelines, but were unable to characterize any of their spectra well enough to apply statistical retrieval techniques.  The most promising young stars have already been searched for substellar companions with high-contrast photometric surveys  \citep{Biller+Liu+Wahhaj+etal_2013,Chauvin+Vigan+Bonnefoy+etal_2014,Brandt+Kuzuhara+McElwain+etal_2014,Bowler+Liu+Shkolnik+etal_2015}.  As a result, additional discoveries by GPI and other high-contrast IFSs will likely be at small separations and high contrasts, where the effects of image processing are hardest to characterize. The same challenges will arise in space-based IFS imaging at exceptionally high contrasts using WFIRST-AFTA \citep{Spergel+Gehrels+Baltay+etal_2015}.

In this paper, we measure the spectral covariance in GPI commissioning and early science data.  Using our measured spectral covariances, and assuming the data reduction to be unbiased, we demonstrate the effects of incorrect error modeling on atmospheric retrievals.  The paper is organized as follows. In Section~\ref{sec:likelihood}, we define the likelihood function. In Section~\ref{sec:spec_cov}, we discuss sources of spectral covariance in IFS data and show how to measure the spectral covariance matrix using actual IFS observations. In Section~\ref{sec:retrievals}, we compare results from mock retrievals that use the full covariance matrix, which we assume to be the source of the noise, with retrievals that set its off-diagonal terms to zero. In Section~\ref{sec:nonuniform}, we explore the additional effect of realistic priors on effective temperature and surface gravity in Bayesian parameter retrievals. We conclude in Section~\ref{sec:conclusion}.

\section{The Likelihood Function} \label{sec:likelihood}

Spectroscopic characterization is the process of inferring an object's physical properties from its measured spectrum.  A spectroscopic observation generally yields a noisy measurement ${\bm S}$ of a source's flux density ${\bm F}$.  In the most general case, the measured spectral values may be biased and coupled by a measurement matrix ${\bm M}$, so that the measured spectrum is given by 
\begin{equation}
    \bm{S} = \bm{M} \cdot \bm{F} + \bm{\varepsilon},
\end{equation}
where ${\bm\varepsilon}$ is the matrix of measurement errors.  Our goal in this paper is to describe the proper treatment of measurement errors; we assume that ${\bm M}$ is known (i.e., its inverse can be applied to ${\bm S}$), and that we can simply write
\begin{equation}
S_i = F_i + \epsilon_i,
\end{equation}
where the index refers to the $i^\mathrm{th}$ wavelength bin, and $\epsilon_i$ is the measurement error associated with that bin.  The value of each $\epsilon_i$ is typically impossible to calculate; spectroscopic characterization relies on knowing its statistical properties.  A forward model maps the physical quantities of interest (temperature, abundances, etc.) onto theoretical spectra, which may be statistically compared to the observed spectrum ${\bm S}$ to derive parameter constraints.

Parameter retrieval begins with the likelihood function ${\cal L}$, the probability of measuring the observed spectrum ${\bm S}$ given a model of the true spectrum ${\bm F}$.  The forward model is a function of a set of physical parameters $\Phi$, making ${\cal L}$ also a function of $\Phi$. If the probability distribution of $\epsilon_i$ in each wavelength bin is Gaussian with zero mean, then the log of the likelihood function is given by 
\begin{equation}\label{eqn:chi2_C}
-2 \ln\mathcal{L}(\Phi) \equiv \chi^2 = \left({\bm S} - {\bm F}\right)^T {\bm C}^{-1}\left({\bm S} - {\bm F}\right),
\end{equation}
where ${\bm C}$ is the covariance matrix with elements
\begin{equation}
C_{ij} \equiv \langle \epsilon_i \epsilon_j \rangle, 
\end{equation}
with $\langle...\rangle$ denoting the expectation value. In the case of independent errors, $C_{ij} = 0$ for $i \neq j$, and Equation~(\ref{eqn:chi2_C}) reduces to the usual 
\begin{equation}
-2 \ln\mathcal{L}(\Phi) = \sum_i \frac{(S_i - F_i)^2}{C_{ii}}.
\label{eqn:chi2_nocov}
\end{equation}

The maximum likelihood estimate is the set of parameters $\Phi$ that maximize ${\cal L}$ or, equivalently, minimize $\chi^2$.  In a Bayesian parameter retrieval, the posterior probability distribution of the parameters $\Phi$ given the observed spectrum ${\bm S}$ is proportional to the product of the likelihood function and the prior probability distribution of $\Phi$:
\begin{equation}
p(\Phi | {\bm S}) \propto {\cal L}(\Phi) p(\Phi).
\label{eq:bayes}
\end{equation}
Equation \eqref{eq:bayes} must then be integrated and normalized to obtain a probability density, for example by Markov Chain Monte Carlo.  Confidence regions are enclosed by surfaces of constant posterior probability density containing a given fraction of the integrated posterior probability.

\begin{figure*}[t!]
\centering
\includegraphics[width=17.5cm]{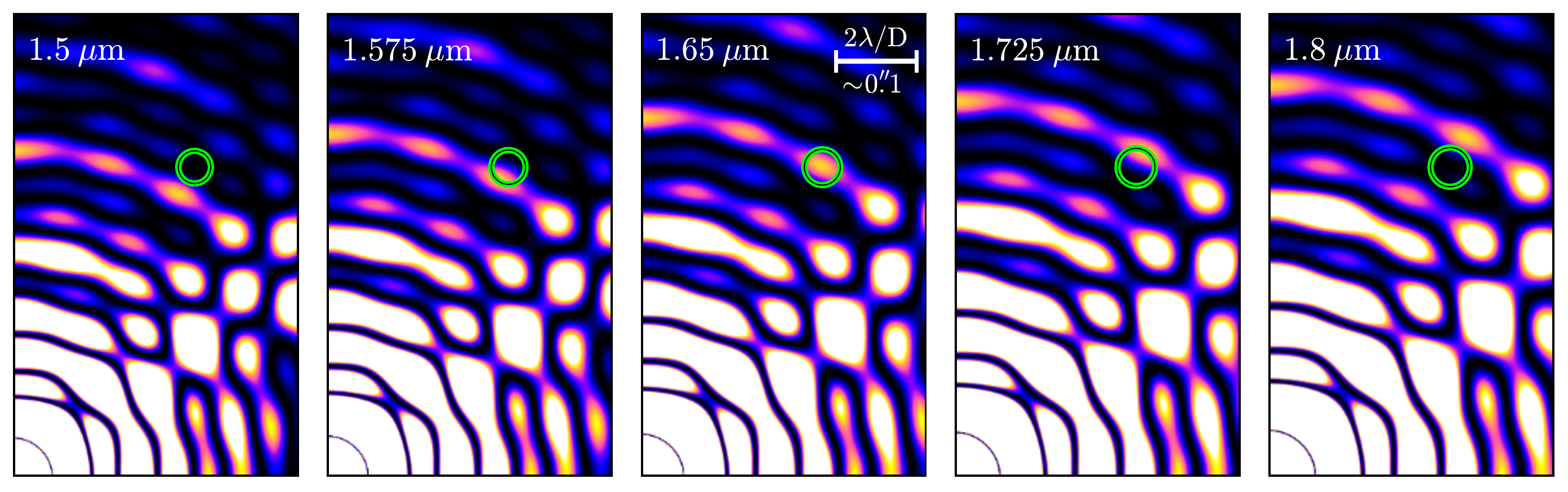}
\caption{A demonstration of the motion of a PSF speckle with respect to the location of a hypothetical planet (green circles). Each frame shows a monochromatic, unocculted PSF with a linear stretch; the center of each PSF defines the lower-left corner of the frame. 
The bandpass width is ${\sim}20\%$, similar to the $H$-band. The image scale is given in units of $\lambda/D$ and arcseconds in the middle frame, assuming $\lambda=1.65~\mu$m and $D=8$~m. The green circles represent a planet located ${\sim}11\,\lambda/D$ from the center of the PSF. The angular diameter of each circle is  $\lambda/D$, the characteristic scale of a planet PSF. The speckle moves approximately two full planet PSF diameters as we step through the ${\sim}20\%$ bandpass, adding flux density within the planet's PSF for roughly half the filter width, which couples neighboring wavelengths at this characteristic spectral resolution.}
\label{fig:toymodel}
\end{figure*}

\section{Spectral Covariance in IFS Data}\label{sec:spec_cov}

In this paper, we explore the importance of using the correct covariance matrix ${\bm C}$ to infer physical parameters from IFS data. Many sources of noise, e.g.~photon noise and read noise, are independent at different wavelengths, allowing Equation~(\ref{eqn:chi2_nocov}) to be used.  The process of reducing high-contrast IFS data, however, introduces sources of noise that are spectrally correlated.  As a trivial example, interpolation to align a series of images or to scale a model PSF can couple neighboring wavelengths.  As another example, the spectrum of each lenslet on the detector is the convolution of the true spectrum and that lenslet's diffraction pattern, with the additional possibility of diffraction or ``crosstalk'' from a neighboring lenslet. Spectral correlation can also be induced by subtracting an imperfect model of the stellar PSF. We explore this effect further with a toy model in Section~\ref{sec:toy}. We then measure spectral covariance in actual IFS data in Section~\ref{sec:parameterize}. 

\subsection{Spectral Covariance from PSF Speckles}\label{sec:toy}

As diffraction phenomena, PSF speckles scale radially with wavelength, moving into and out of a planet's location. This scaling may be exploited to suppress speckles in high-contrast images \citep{Racine+Walker+Nadeau+etal_1999, Marois+Doyon+Racine+etal_2000, Sparks+Ford_2002}. Imperfectly subtracted speckles will add flux density to a planet's spectrum with a characteristic spectral resolution, which depends upon the angular separation between the planet and star in units of $\lambda/D$, where $D$ is the effective telescope diameter. For a change in wavelength $\Delta\lambda$, a speckle's angular distance from the star $\rho$ will vary by $\Delta\rho = \rho\,\Delta\lambda/\lambda$. Thus, for a fixed bandpass width $\Delta \lambda/\lambda$, the rate (with respect to a change in wavelength) at which a speckle moves across the core of a planet's PSF will be proportional to its angular separation from the star. This effect introduces a characteristic spectral correlation length that scales linearly with angular separation.

As an intuitive demonstration of how PSF speckles can induce spectral correlation, we construct a toy model that consists of a series of monochromatic, unocculted PSFs spanning a bandpass of width $\Delta\lambda/\lambda\sim20\%$ (similar to the $H$-band). For each wavelength, we generate a PSF by taking the Fourier transform of a uniformly illuminated pupil and squaring the amplitude to get the intensity. We then observe a speckle as it moves across the location of a hypothetical planet. The actual pupil we use, as well as our model's lack of a coronagraph, are not important to us here; the radial scaling of speckles is independent of these details. 

Figure~\ref{fig:toymodel} shows the expansion of the PSF as we step evenly across the ${\sim}$20\% bandpass in five wavelength frames. The indicated wavelengths span the $H$-band, and the angular scale given in the middle frame assumes an 8~m primary. The green circles represent a possible location of a planet. Each circle has an angular diameter of $\lambda/D$ and is located ${\sim}11\,\lambda/D$ from the center of the PSF, which defines the lower-left corner of each frame. At this separation, the radial position of a speckle changes by $\Delta \rho \sim 0.2\times11\,\lambda/D\sim2\lambda/D$ (i.e., 2 planet PSF diameters) across the bandpass. The speckle will add flux density within the planet's PSF for approximately half the filter width, coupling neighboring wavelengths at this characteristic spectral resolution. 

\subsection{Measuring the Spectral Covariance Matrix} 
\label{sec:parameterize}

\begin{figure*}[t!]
\centering
\includegraphics[width=17.6cm]{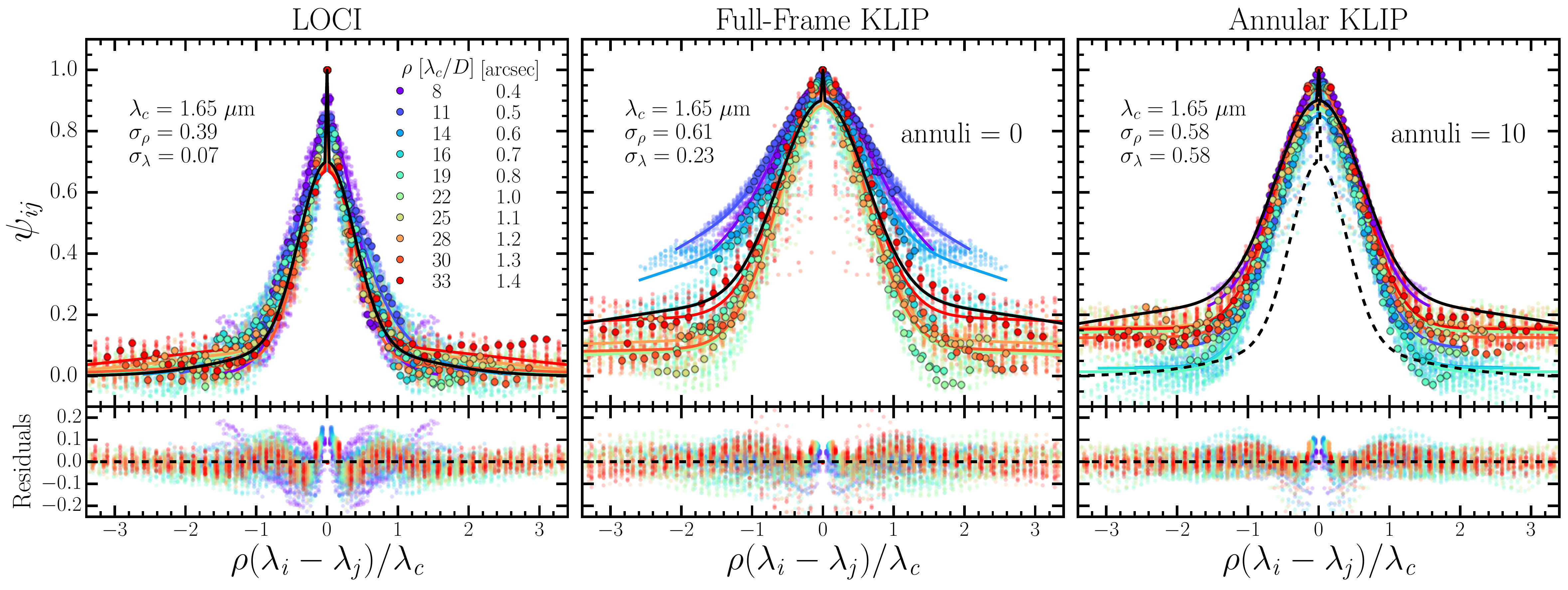}
\caption{The measured spectral correlation $\psi_{ij}$, as a function of angular separation $\rho$, in the $H$-band data cube of HIP~21861, where we have used LOCI (left panel), full-frame KLIP (annuli~=~0; middle panel), and KLIP with 10 annuli (right panel) for PSF subtraction. Motivated by the $\rho$-dependence of PSF-subtraction-induced spectral correlation (see Section~\ref{sec:toy}), we plot the correlation functions against $\rho(\lambda_i - \lambda_j)/\lambda_c$, where $\lambda_c$ is the central wavelength of the bandpass. The colors in each panel correspond to the separations indicated in the left panel in units of $\lambda_c/D$ and arcseconds (assuming $D=8$~m). Small dots show the measured correlation for all possible pairs ($i$, $j$) of wavelengths, and the larger circles show the correlation for $\lambda_i = \lambda_c = 1.65~\mu$m (i.e., the middle row of the correlation matrix). The colored lines show the best fits to Equation~(\ref{eqn:fit_psi}) for each separation, and the lower panels show the associated residuals. Black lines show our fiducial model of the spectral correlation with our LOCI and KLIP noise parameters, which we adopt to perform mock retrievals in Section~\ref{sec:retrievals}. In the right panel, we show our fiducial models for both KLIP (solid) and LOCI (dashed). LOCI is the most local of these PSF subtraction algorithms and full-frame KLIP is the most global; they bracket a plausible parameter space of spectral covariance.}
\label{fig:psi}
\end{figure*}

Even in a multi-planet image, most of the field-of-view is devoid of planets. If no other astrophysical sources are present (e.g., a disk), spectra of these empty regions are realizations of the spectral noise and can be used to measure both its amplitude and covariance. Here, we demonstrate such measurements using reduced data cubes from the GPI collaboration's First Data Release\footnote{http://www.gemini.edu/sciops/instruments/gpi/public-data}. We start with the data cubes extracted by the GPI team from the two-dimensional detector readouts and use the GPI data reduction pipeline \citep{Maire+Perrin+etal_2010, Perrin+Maire+Ingraham+etal_2014} to implement the LOCI \citep[][GPI primitive: ``ADI with LOCI'']{Lafreniere+Marois+Doyon+etal_2007} and KLIP \citep[][GPI primitive: ``KLIP algorithm Angular Differential Imaging'']{Soummer+Pueyo+Larkin_2012} algorithms for PSF subtraction. Both algorithms operate separately on each wavelength and model the PSF with a number of basis images using least-squares optimization; they differ (in GPI's implementation) in {\it where} the approximation is computed.  LOCI approximates the PSF locally in radially extended wedges, while KLIP performs a separate approximation in a user-supplied number of annuli.  With annuli~=~0 (the default in the GPI pipeline), KLIP performs a global approximation to the PSF.  We adopt LOCI and full-frame (annuli~=~0) KLIP as our fidicial reductions.  We also compute the spectral covariance for KLIP with annuli~=~10; this is more local than full-frame KLIP but less local than LOCI. All other parameters are fixed at GPI's default values (LOCI parameters: nfwhm~=~2.5, coeff\_type~=~0; KLIP parameters: MinRotation~=~1$^\circ$, prop~=~.99999). 

Throughout, we assume the data reduction to be unbiased; for example, the measurement errors are independent of a planet's spectrum. We stress that it is not our goal to fully characterize the noise in GPI data. We instead present a practical approach to dealing with correlated noise in IFS observations and study the general consequences of ignoring spectral covariance when estimating atmospheric parameters from recovered exoplanet spectra. The PSF-subtracted ($H$-band) data cube of HIP~21861 is by far the cleanest of all the targets in this data release. Since we expect the quality of IFS data will only improve with time, we choose to adopt the HIP~21861 data set as representative of this class of observations. Table \ref{tab:gpi_data_set} summarizes the properties of this observation.

\begin{deluxetable}{lr}
\tablewidth{0pt}
\tablecaption{The HIP 21861 GPI Data Set}
\tablehead{
    Parameter & 
    Value
    }
\startdata
Observation Date & 14 November 2013 \\
Target $V$-band Brightness & 5.0 mag\tablenotemark{*} \\
Total Time on Target & 70~minutes \\
Total Exposure Time & 60~minutes \\
Field Rotation & $95^\circ$ \\
Mean Airmass & 1.01 \\
Seeing Range & $0.\!\!''5$--$0.\!\!''7$
\enddata
\tablenotetext{*}{From the Tycho-2 catalog \citep{Hog+Fabricius+Makarov+etal_2000}}
\label{tab:gpi_data_set}
\end{deluxetable}

Given a reduced, PSF-subtracted data cube, we measure the average spectral correlation at fixed angular separation $\rho$ from the central star as follows. Assuming azimuthal symmetry, we extract a host-star-centric annulus of radius $\rho$ from each wavelength frame of the data cube. The thickness of the annulus is taken to be 4 pixels (we varied this thickness by ${\pm50\%}$ and found that it does not significantly change our results). We then compute the average spectral correlation within the annulus,
\begin{equation}\label{eqn:psi}
\psi_{ij} \equiv \frac{C_{ij}}{\sqrt{C_{ii}C_{jj}}} = \frac{\langle I_i I_j \rangle}{\sqrt{\langle I_i^2 \rangle \langle I_j^2 \rangle}},
\end{equation}
where $I_i$ is the intensity at wavelength $i$, and $\langle ... \rangle$ is the expectation value over all spatial locations within the annulus. Uncorrelated (white) noise would have $\psi_{ij}=0$ for $i \neq j$. 

Figure~\ref{fig:psi} shows the measured spectral correlation in the $H$-band data cube of HIP~21861, where we have used LOCI (left panel), full-frame KLIP (annuli~=~0; middle panel), and KLIP with 10 annuli (right panel) for PSF subtraction. We plot the correlation functions against $\rho(\lambda_i - \lambda_j)/\lambda_c$, where $\lambda_c$ is the central wavelength of the bandpass, and $\rho$ is in units of $\lambda_c/D$ with $D=8$~m for Gemini. This scaling is motivated by the characteristic correlation length induced by imperfect PSF subtraction, which scales with angular distance from the central star (see Section~\ref{sec:toy}). The colors in each panel correspond to the separations indicated in the left panel in units of $\lambda_c/D$ and arcseconds. The small dots show the measured correlation for all possible pairs ($i$, $j$) of wavelengths, and the larger circles show the correlation for $\lambda_i = \lambda_c = 1.65~\mu$m (i.e., the middle row of the correlation matrix). All panels show spectral correlation on several characteristic scales. In addition, the correlation functions for both LOCI and KLIP show a sharp peak at the origin, suggesting a significant component of the noise is uncorrelated. 

We parameterize the spectral correlation $\psi_{ij}$ with a multi-component noise model consisting of two correlated noise terms and an independent noise term:
\begin{align}\label{eqn:fit_psi}
\psi_{ij} \approx &A_\rho \exp\left[-\frac{1}{2}\left(\frac{\rho}{ \sigma_\rho}\frac{\lambda_i - \lambda_j}{\lambda_c}\right)^2\right] \nonumber \\ &+ A_\lambda \exp\left[-\frac{1}{2}\left(\frac{1}{ \sigma_\lambda}\frac{\lambda_i - \lambda_j}{\lambda_c}\right)^2\right] + A_\delta\,\delta_{ij},
\end{align}
where $A_\rho$ and $A_\lambda$ are the amplitudes of the correlated noise terms, $A_\delta$ is the amplitude of the independent noise term, $\sigma_\rho$ and $\sigma_\lambda$  are correlation lengths that characterize the corresponding noise terms, and $\delta_{ij}$ is the Kronecker delta. The two correlated noise terms model spectral correlations at different scales. The first term scales with angular separation and models correlations induced from incomplete PSF subtraction. The second term is independent of angular separation and models other sources of spectral correlation, such as interpolation or spectral crosstalk, which do not scale with angular distance from the central star. The Kronecker delta only contributes to the diagonal of the correlation matrix, allowing a component of the noise to be uncorrelated. When fitting for these parameters, we allow the three amplitudes to vary with angular separation but force $\sigma_\rho$ and $\sigma_\lambda$ to be constant across the field-of-view. In an actual parameter retrieval, it may be more appropriate to restrict the fits to a few separations near the planet's location rather than the entire field-of-view, but our aim here is to gain some intuition for the problem and identify potential sources of spectral covariance. In addition, we require the amplitudes to sum to unity. The colored lines in Figure~\ref{fig:psi} show the best fits for each separation, and the lower panels show the associated residuals.

We note that, given the two correlation lengths ($\sigma_\rho$ and $\sigma_\lambda$), fitting for the amplitudes is a linear problem. Therefore, it is computationally inexpensive to find the best-fit amplitudes; in this case, sampling algorithms such as Markov Chain Monte Carlo are not necessary. By calculating the best-fit parameters over a sufficiently large (two-dimensional) grid of correlation lengths, one can be certain that they have found the global minimum of the fitting function.

For LOCI, we find typical amplitudes of $A_\rho\sim0.6$, $A_\lambda\sim0.1$, and $A_\delta\sim0.3$, with best-fit correlation lengths $\sigma_\rho = 0.39$ and $\sigma_\lambda = 0.07$. For full-frame KLIP, we find typical amplitudes of $A_\rho\sim0.65$, $A_\lambda\sim0.25$, and $A_\delta\sim0.1$, with best-fit correlation lengths $\sigma_\rho = 0.61$ and $\sigma_\lambda = 0.23$. We refer to these sets of parameters as our LOCI and KLIP noise parameters, respectively, and adopt them as our fiducial model of the spectral correlation in the following sections. The black lines in Figure~\ref{fig:psi} show these parameterizations of the spectral correlation, which combine with Equation~(\ref{eqn:psi}) to provide a realistic model of the covariance matrix ${\bm C}$. For comparison, KLIP with 10 annuli produces typical amplitudes of $A_\rho\sim0.8$, $A_\lambda\sim0.1$, and $A_\delta\sim0.1$, with best-fit correlation lengths $\sigma_\rho = 0.58$ and $\sigma_\lambda = 0.58$. In the right panel of Figure~\ref{fig:psi}, the black lines show the correlation function assuming our fiducial KLIP (solid) and LOCI (dashed) parameters defined above. Generally speaking, the spectral covariance for KLIP with annuli~=~10 is intermediate between that for LOCI and for full-frame KLIP.

The spectral correlations for both LOCI and KLIP have a dominant ($A_\rho\gtrsim60\%$) noise component that is correlated on a scale of $\sigma_\rho\sim0.5$ in units of $\lambda_c/D$. This is consistent with our expectation from Section~\ref{sec:toy} of spectral correlations induced by imperfect PSF subtraction. Both LOCI and KLIP also have a significant component of uncorrelated noise, with LOCI having more power in this term. The most significant difference between LOCI and full-frame KLIP is associated with the second noise term in Equation~(\ref{eqn:fit_psi}); KLIP produces much stronger large-scale correlations, whereas LOCI appears to be more sensitive to small-scale coupling between neighboring wavelengths.

LOCI and KLIP are mathematically similar and, in GPI's implementation, both operate separately on each wavelength. The reasons for the striking differences in their covariance matrices are subtle. LOCI calculates its PSF approximation locally within radially extended wedges; this radial extent means that LOCI has some knowledge of speckles that, as shown in Figure~\ref{fig:toymodel}, enter the subtraction region at shorter wavelengths.  In contrast, GPI's implementation of KLIP is either a global (annuli~=~0) or strictly annular algorithm. At a given wavelength, there is less effect from speckles that enter the subtraction region at other wavelengths. Global fits may also have larger residuals due to the need to fit regions of high intensity, which are disproportionately weighted in least squares. These residuals may have power on many spatial scales. Implementations of KLIP using different geometries, for example radially extended wedges or with smaller annuli, might produce very different spectral covariances. This highlights that the spectral covariance is highly sensitive to the details of how the PSF subtraction algorithm is implemented. It is essential to measure the spectral covariance for each data set and reduction technique. In the GPI pipeline, LOCI and full-frame KLIP (annuli~=~0) generate spectral covariance using the most local and global PSF approximations, respectively; we use these for our examples in the rest of this paper.

\section{Impact on Inferred Parameters}\label{sec:retrievals}

We now combine our realistic noise model with theoretical spectra to investigate the effect of spectral covariance on atmospheric parameter retrievals. Assuming the solar-metallicity BT-Settl spectral models \citep{Allard+Homeier+Freytag_2011} as ``truth'', we select a particular spectrum, add a realization of noise using our measured spectral errors and covariances, and attempt to recover the model's input effective temperature ($T_\mathrm{eff}$) and surface gravity ($\log g$). Only by guaranteeing that the true spectrum is in our fitting library may we satisfy the assumptions of Section~\ref{sec:likelihood}. We perform thousands of parameter retrievals with and without accounting for the full covariance matrix that we assume to be the source of the noise (i.e., using the full covariance matrix and setting the off-diagonal terms to zero). We run our analysis using both our LOCI and KLIP noise parameterizations. In Section~\ref{sec:uniform}, we carry out the retrievals with uniform priors in $T_\mathrm{eff}$ and $\log g$. In Section~\ref{sec:nonuniform}, we study the additional effect of realistic priors in Bayesian parameter retrievals. 

\begin{figure}[t!]
\centering
\includegraphics[width=8.cm]{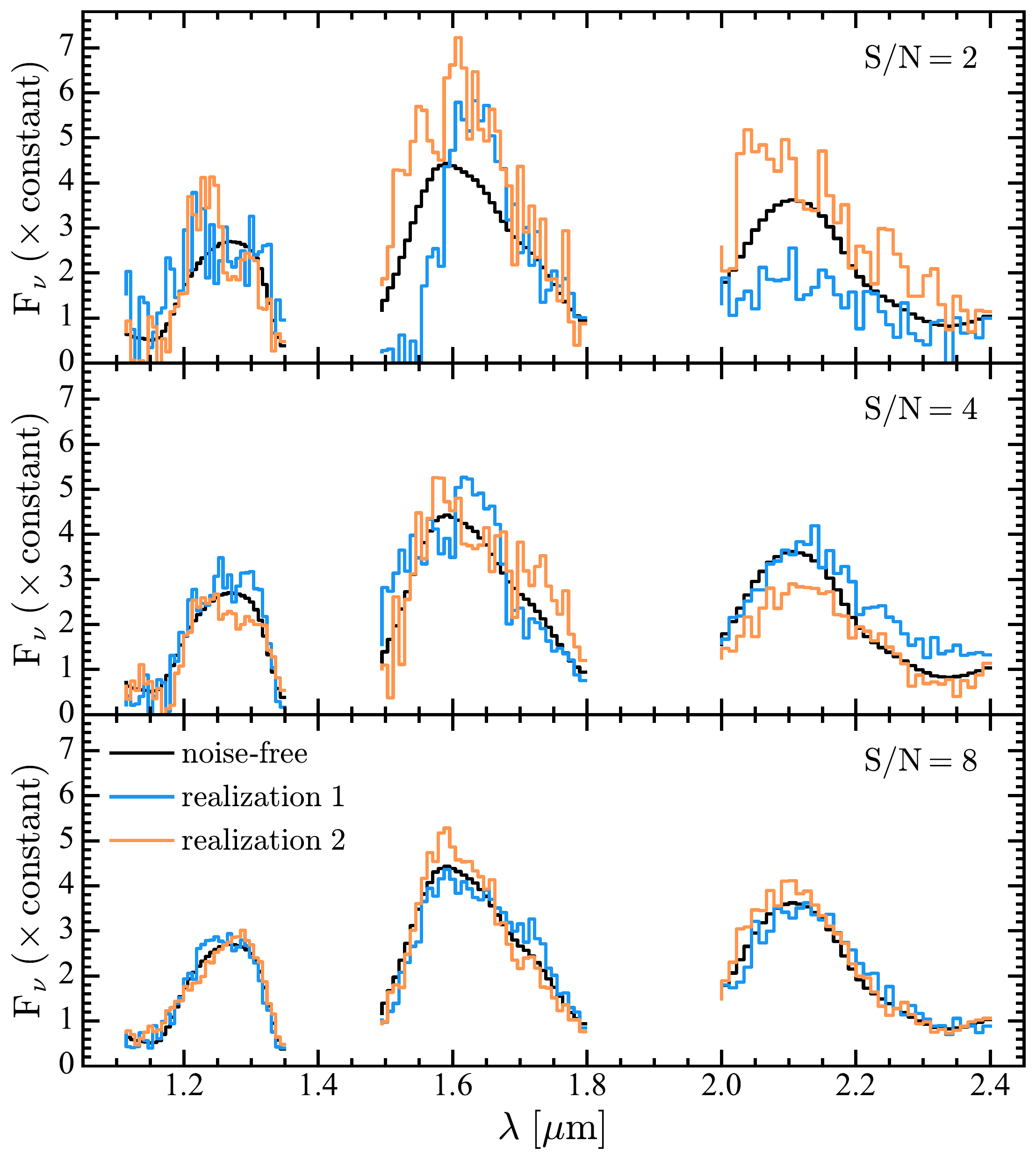}
\caption{Mock spectra of a planet with $T_\mathrm{eff}=1000$, $\log g=4.0$, and solar metallicity generated from the BT-Settl spectral models \citep{Allard+Homeier+Freytag_2011}. All spectra are smoothed to a constant resolution of $\lambda/\Delta\lambda=45$ using a Gaussian line spread function and resampled with 37 wavelength bins in $J$, $H$, and $K$. The black lines show the noise-free spectrum of the planet, and the blue and orange lines show different realizations of the spectrum with correlated Gaussian noise (generated with our LOCI parameterization), assuming a mean S/N of 2 (top), 4 (middle), and 8 (bottom). For each band, we define the mean S/N to be the mean flux across the band divided by the root-mean-square of the noise.}
\label{fig:noisy_data}
\end{figure}

\begin{figure*}[t!]
\centering
\includegraphics[width=16.8cm]{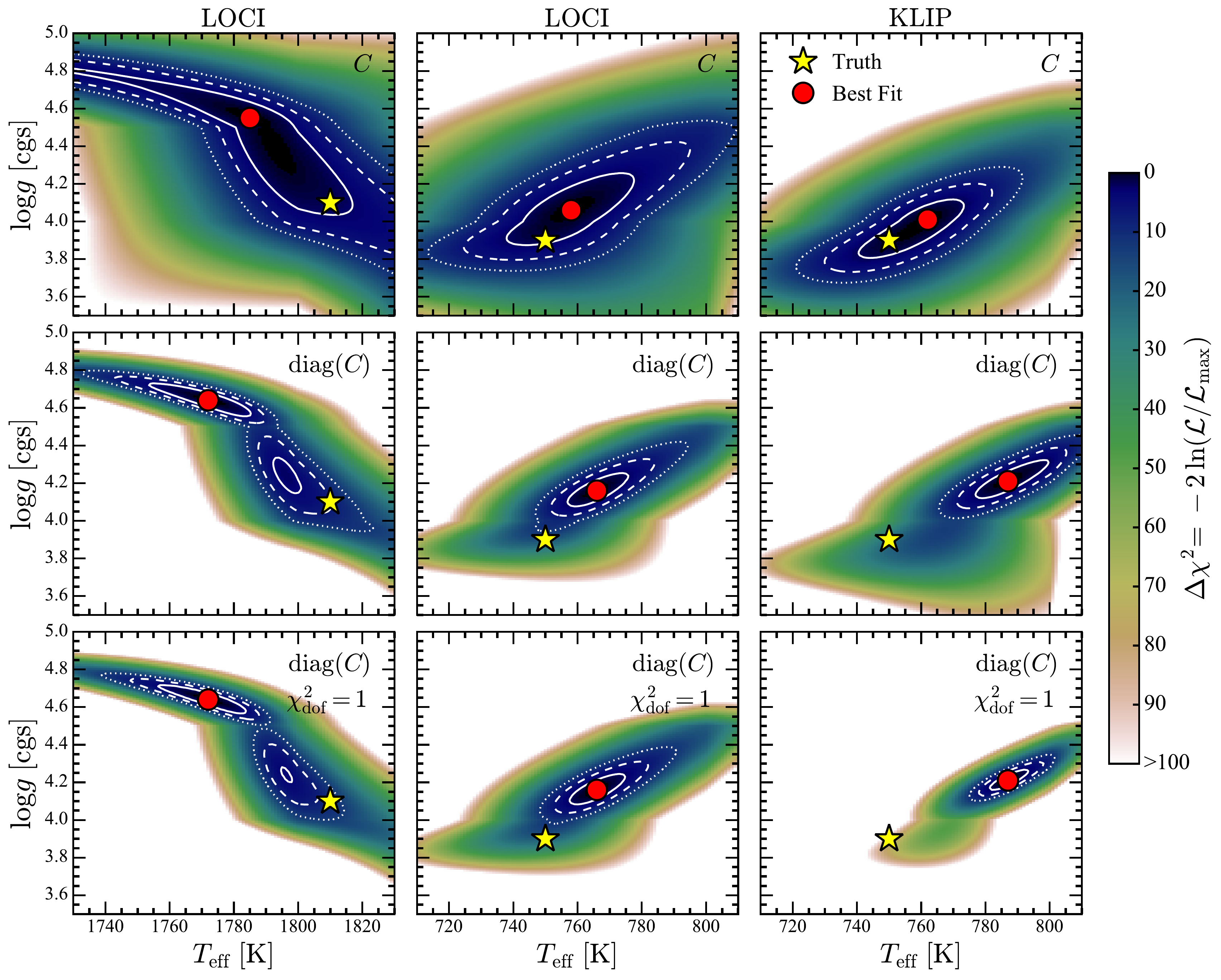}
\caption{Representative grids of $\Delta \chi^2 \equiv \chi^2 - \chi^2_\mathrm{min} = -2 \ln\left(\mathcal{L}/\mathcal{L}_\mathrm{max}\right)$ for a high-temperature ($T_\mathrm{eff}=1810$~K; left column) and low-temperature ($T_\mathrm{eff}=750$~K; middle column) input spectrum, assuming our LOCI noise parameterization. The right column shows the low-temperature case assuming our KLIP noise parameterization. In the top row, we calculate $\chi^2$ using the full covariance matrix ($C$ case; Equation~(\ref{eqn:chi2_C})), and in the middle and bottom rows, we use the diagonal of the covariance matrix (diag($C$) case; Equation~(\ref{eqn:chi2_nocov})). Yellow stars indicate the model's input parameter values, and red circles indicate the best-fit parameters. The solid, dashed, and dotted lines show the 68\%, 95\%, and 99\% confidence regions, respectively, which are determined by integrating $\mathcal{L}$. In the bottom row, we show the same $\Delta \chi^2$ grids as in the middle row, except rescaled so that the $\chi^2$ per degree of freedom ($\chi^2_\mathrm{dof}$) is unity. We used the same realization of noise to generate each column; differences between the top and middle rows are due solely to whether or not spectral covariance was included in the fitting procedure.}
\label{fig:dchi2}
\end{figure*}

\subsection{Mock Spectra Generation}\label{sec:mock}

We generate mock spectra from the BT-Settl spectral models; these are provided on grids of effective temperature, surface gravity, and metallicity. We interpolate between models using piecewise power laws and smooth each spectrum to a constant resolution of $\lambda/\Delta\lambda=45$ using a Gaussian line spread function with full width at half maximum $\mathrm{FWHM}=\Delta \lambda/\lambda$. We then resample in $J$, $H$, and $K$ with 37 wavelength bins per band, which is similar to the wavelength sampling in the GPI data cubes and somewhat higher than the Nyquist sampling rate. Finally, to each mock spectrum, we add Gaussian noise with spectral covariance given by our LOCI and KLIP noise parameterizations. For the variances $C_{ii}$, we use the measured variances at a separation of $0.\!\!''7\sim 16\,\lambda_c/D$ in the $H$-band data cube of HIP~21861. Although we only have measurements in $H(1.65~\mu$m, $16\,\lambda_c/D$), we use the same (wavelength-dependent) noise parameterizations for $J(1.23~\mu$m, $22\,\lambda_c/D$) and $K(2.1~\mu$m, $12\,\lambda_c/D$). We note that it is likely that measurements in each band would produce somewhat different noise parameterizations. In an actual parameter retrieval, the covariance matrix should be measured in each band.

Figure~\ref{fig:noisy_data} shows the noise-free spectrum (black lines) of a planet with $T_\mathrm{eff}=1000~K$, $\log g = 4.0$, and solar metallicity along with two realizations of the spectrum with correlated Gaussian noise (generated with our LOCI noise parameterization),  assuming a mean signal-to-noise (S/N) of 2 (top panel), 4 (middle panel), and 8 (bottom panel) in each band. Here, we define S/N to be the mean flux across each band divided by the root-mean-square of the noise. Correlated noise can significantly alter the shape of broad features (e.g., the triangular shape in the $H$-band), which can impact parameters inferred from such data. 

\subsection{Inferring $T_\mathrm{eff}$ and $\log g$} \label{sec:uniform}

\begin{figure}[t!]
\centering
\includegraphics[width=8.4cm]{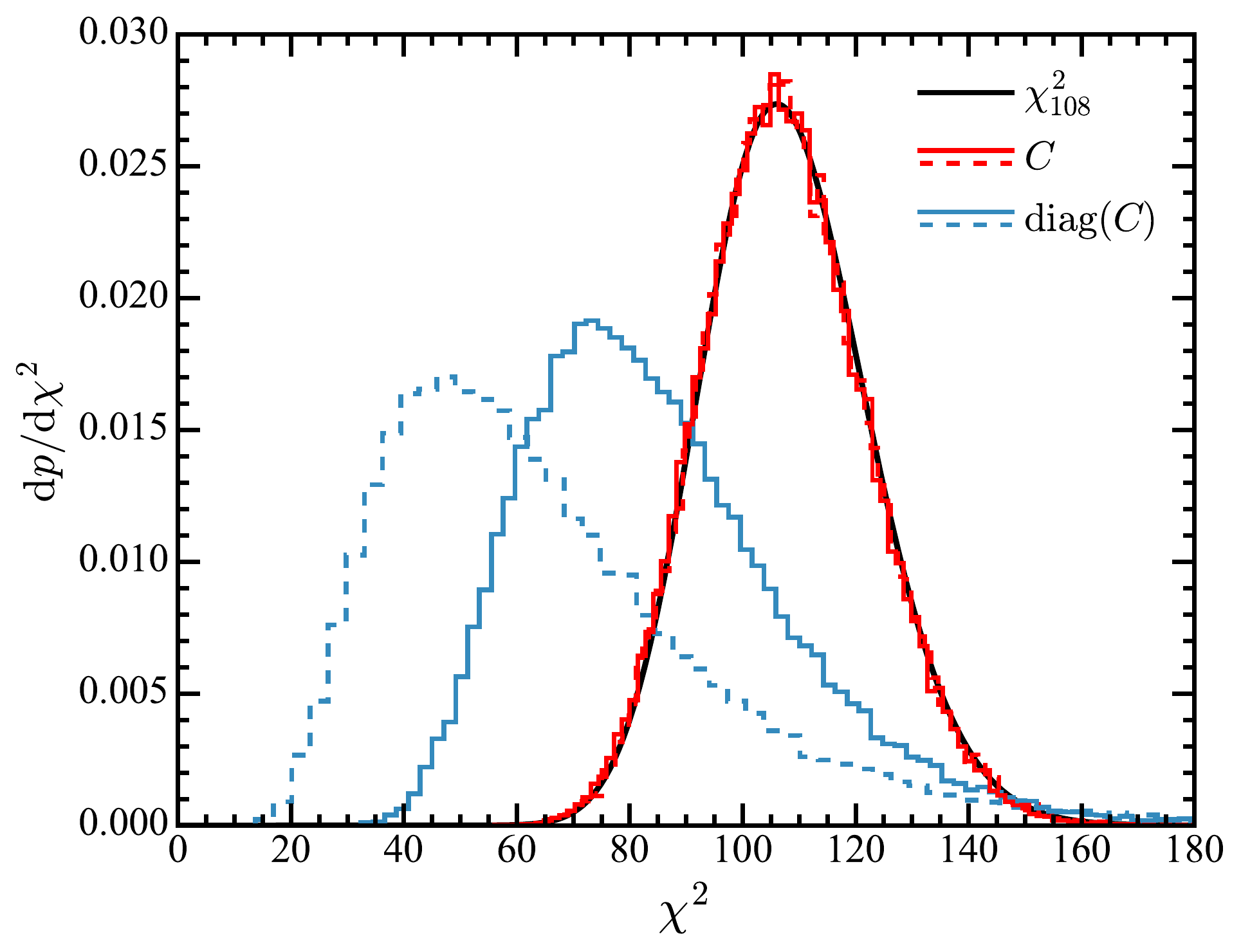}
\caption{$\chi^2$ distributions from $5\times10^4$ retrievals of $T_\mathrm{eff}$ and $\log g$, where we calculate $\chi^2$ using the full covariance matrix (red histograms; Equation~(\ref{eqn:chi2_C})) and the diagonal of the covariance matrix (blue histograms; Equation~(\ref{eqn:chi2_nocov})). The solid (dashed) lines show the distributions assuming our LOCI (KLIP) noise parameters. The solid black line shows the theoretical expectation for $\chi^2$ with Gaussian noise and 108 degrees of freedom $\left(\chi^2_{108}\right)$. As expected, we recover the appropriate $\chi^2$ distribution with the correct error model. However, ignoring spectral covariance tends to produce $\chi^2<108$, the expectation value of $\chi^2_{108}$, with the offset being more pronounced for KLIP.}
\label{fig:chi2_hist}
\end{figure}

To explore how spectral covariance influences parameter estimation, we perform thousands of retrievals of $T_\mathrm{eff}$ and $\log g$ using the standard method of $\chi^2$-minimization on a grid of forward models, which is widely used in the literature \citep[e.g.,][]{Saumon+Marley+etal_2006,Cushing+Marley+etal_2008,Rice+Barman+etal_2010, Madhu+Burrows+Currie_2011, Bonnefoy+Currie+Marleau+etal_2014, Rice+Oppenheimer+Zimmerman+etal_2015}. Our approach may be naturally extended to the inverse methods for atmospheric retrievals that are actively being developed for exoplanetary studies \citep[e.g.,][]{Line+Wolf+etal_2013, Lee+Heng+Irwin_2013, Line+Teske+etal_2015}. In all our calculations, we assume solar metallicity.

For each retrieval, we generate a mock spectrum as described in Section \ref{sec:mock}, adding a realization of spectral noise with S/N=10 to a BT-Settl model. We then build grids of $\chi^2 = -2 \ln \mathcal{L}$ using both Equations~(\ref{eqn:chi2_C}) and (\ref{eqn:chi2_nocov}). We define the former to be the ``$C$ case'', since the retrieval is performed with the full covariance matrix, and the latter to be the ``diag$(C)$ case'', since the off-diagonal elements of the covariance matrix are set to zero. The best-fit parameters are then given by the maximum likelihood or, equivalently, the minimum $\chi^2$. For each $\left(T_\mathrm{eff},\ \log g\right)$ pair, we allow the normalization of the spectrum to vary as an additional free parameter, resulting in $37\times3 - 3 = 108$ degrees of freedom. Allowing the normalization to float accounts for the generic difficulty models have with producing the correct luminosity in $J$, $H$, and $K$. 

Figure~\ref{fig:dchi2} shows example grids of $\Delta \chi^2 \equiv \chi^2 - \chi^2_\mathrm{min} = -2 \ln\left(\mathcal{L}/\mathcal{L}_\mathrm{max}\right)$ in the $T_\mathrm{eff}-\log g$ plane for a high-temperature ($T_\mathrm{eff}=1810$~K; left column) and low-temperature ($T_\mathrm{eff}=750$~K; middle column) input spectrum, where we have modeled the noise with our fit to the LOCI spectral covariance. In the right column, we show the low-temperature case using our KLIP noise parameters. In the top row, we include spectral covariance in our calculations of $\Delta \chi^2$; we ignore it in the middle and bottom rows. To generate each column, we use a single realization of noise; differences between the top and middle rows are due solely to whether or not spectral covariance was included in the fitting procedure. The input model's parameter values are indicated in each panel by yellow stars, and the best-fit parameters are shown as red circles. The solid, dashed, and dotted lines show the 68\%, 95\%, and 99\% confidence regions, respectively. These regions are determined by integrating the likelihood function, which is equivalent to a Bayesian analysis with flat priors in $T_\mathrm{eff}$ and $\log g$. In the bottom row, we show the same $\Delta \chi^2$ grids as the middle row, except rescaled so that the $\chi^2$ per degree of freedom ($\chi^2_\mathrm{dof}$) is unity. In all but one case shown in Figure~\ref{fig:dchi2}, ignoring spectral covariance leads to the true parameter value falling outside the 95\% confidence region. Furthermore, rescaling so that $\chi^2_\mathrm{dof}=1$ shrinks the confidence regions, which exacerbates the problem. 

Although this figure only shows results for a few realizations of noise, it is representative of the differences we generally see between retrievals that ignore spectral covariance and those that account for it. For both our LOCI and KLIP noise parameters, retrievals that ignore covariance tend to produce confidence regions that are tighter than those that include it. More importantly, the confidence regions in retrievals that ignore covariance are unreliable. In $5\times10^4$ realizations, the ``true'' parameter values in such retrievals fell within the 95\% confidence regions only ${\sim}33\%$ of the time for LOCI and only ${\sim}23\%$ of the time for KLIP. When $\chi^2_\mathrm{dof}$ is set to unity, these fractions become ${\sim26}\%$ and ${\sim}14\%$, respectively. The biases in the inferred parameters are generally smaller than the claimed uncertainties for many substellar companions, but are larger than the uncertainties in our simulated retrievals.  As uncertainties and systematics in spectral modeling fall, systematics in retrieval techniques will become increasingly important.

In Figure~\ref{fig:chi2_hist}, we show the $\chi^2$ distributions from $5\times10^4$ retrievals of $T_\mathrm{eff}$ and $\log g$, where we use the full covariance matrix (red histograms) and set its off-diagonal terms to zero (blue histograms). The solid (dashed) lines show the distributions assuming our LOCI (KLIP) noise parameters. The solid black line shows the theoretical expectation for $\chi^2$ with Gaussian noise and 108 degrees of freedom $\left(\chi^2_{108}\right)$. As expected, we recover the appropriate $\chi^2$ distribution with the correct error model ($C$ case). Ignoring the off-diagonal elements of the covariance matrix (diag($C$) case), however, tends to yield $\chi^2$ values that are systematically lower, with the offset being more pronounced for our KLIP noise parameters. This result has important implications, as it means that simply rescaling the uncertainties so that $\chi^2_\mathrm{dof}=1$ will generally shrink the confidence regions, making the true parameter values fall outside the 95\% contours $75\%-85\%$ of the time (c.f. bottom row of Figure~\ref{fig:dchi2}).

\section{The Effect of Realistic Priors}\label{sec:nonuniform}
\begin{figure}[h!]
\centering
\includegraphics[width=8.14cm]{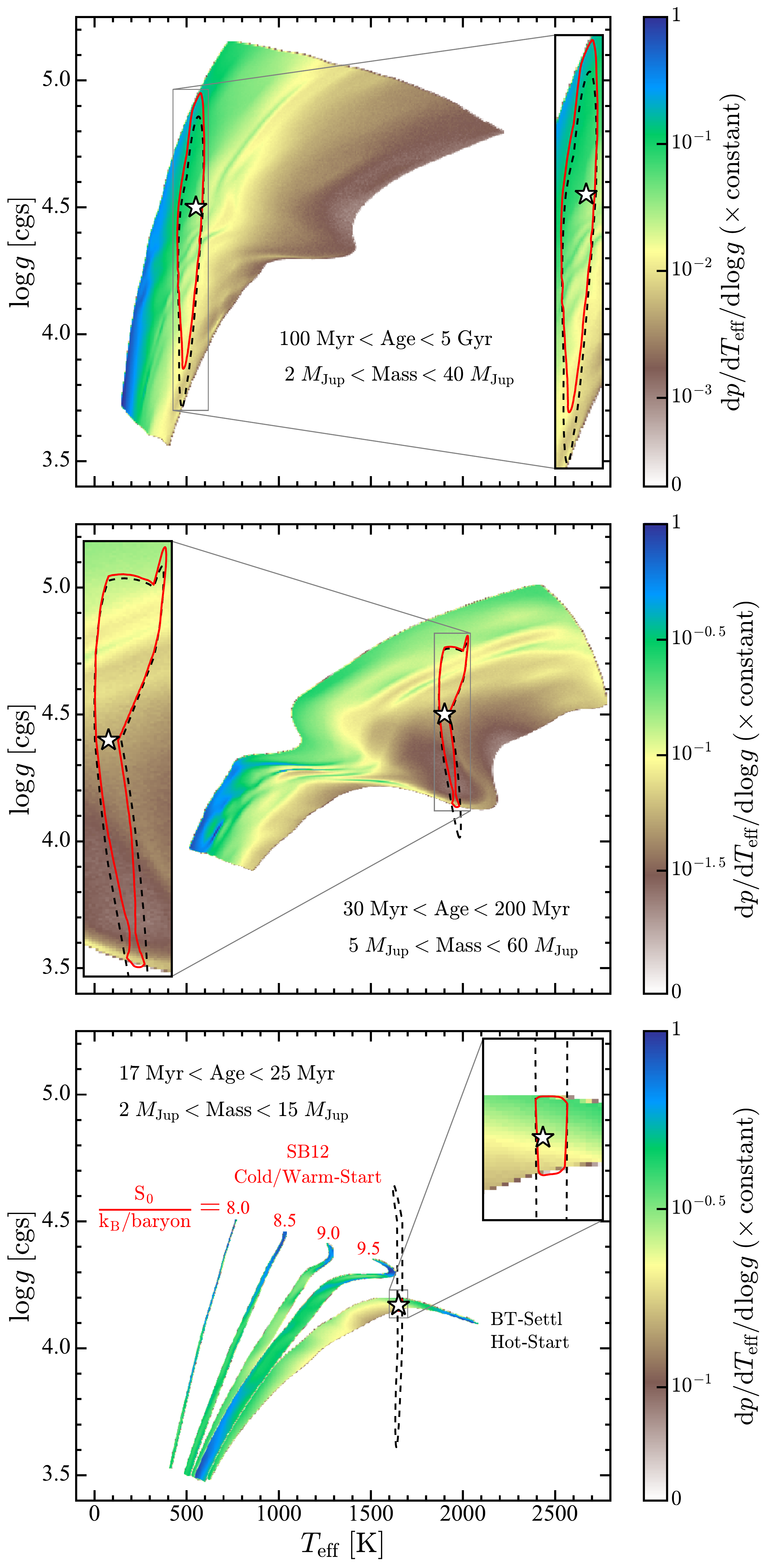}
\caption{Prior probability distributions, $dp/dT_\mathrm{eff}/d\log g$, assuming flat priors on the age and logarithmic mass distributions of substellar companions. For the substellar cooling model, we use the solar-metallicity BT-Settl models \citep{Baraffe+Chabrier+Barman+etal_2003, Allard+Homeier+Freytag_2011}. The indicated age and mass ranges were chosen to be representative of current estimates for GJ~504b (top panel), $\kappa$ And b (middle panel), and $\beta$ Pic b (bottom panel). We perform mock parameter retrievals on the $T_\mathrm{eff}$ and $\log g$ values indicated by the white stars in each panel. The dashed black and solid red contours show the 95\% confidence regions associated with each retrieval, where we have ignored and included the effect of the prior, respectively. The insets magnify where the contours and priors overlap. Each distribution has been normalized such that its peak is unity. In the bottom panel, we also show priors for a series of \citet[][SB12]{Spiegel+Burrows_2012} models with the indicated initial entropies ($S_0$) in units of Boltzmann's constant (k$_\mathrm{B}$) per baryon. When the age is well-constrained, the prior is dominated by evolutionary model uncertainties.}
\label{fig:prior}
\end{figure}

In the parameter retrievals shown in Figure~\ref{fig:dchi2}, we simply mapped the full shape of the likelihood function. This approach is equivalent to a Bayesian analysis when $T_\mathrm{eff}$ and $\log g$ have uniform priors. However, given assumptions about the mass and age distributions of substellar companions and a substellar cooling model, the prior probability distribution in the $T_\mathrm{eff}-\log g$ plane $\left(dp/d\log g/dT_\mathrm{eff}\right)$ will not, in general, be uniform. Here, we make simple assumptions to calculate a plausible prior $dp/d\log g/dT_\mathrm{eff}$, showing the effect nonuniform priors can have on recovered $T_\mathrm{eff}$ and $\log g$ values. 

We start by assuming flat priors on the age and logarithmic mass distributions of substellar companions: $dp/dt/d\log M\sim\mathrm{constant}$. Our assumption about the mass distribution is roughly consistent with current observations \citep[e.g.,][]{Cumming+Butler+Marcy+etal_2008, Brandt+McElwain+Turner+etal_2014}. The flat prior on age is representative of the common observational scenario in which upper and lower limits on the system's age are all that is known. In this simplified picture, the prior is given by the Jacobian:
\begin{equation}\label{eqn:jacobian}
\frac{dp}{dT_\mathrm{eff}\,d\log g} \propto \left| \frac{d\log M}{dT_\mathrm{eff}}\frac{dt}{d\log g} - \frac{d\log M}{d\log g}\frac{dt}{dT_\mathrm{eff}} \right|.
\end{equation}
To calculate $dp/dT_\mathrm{eff}/d\log g$, one can either use Equation~(\ref{eqn:jacobian}) directly or Monte Carlo sample $dp/dt/\log M$ and change coordinates into the $T_\mathrm{eff}-\log g$ plane. We choose the latter, as it avoids numerical issues associated with singularities in the calculation of the Jacobian. We use the solar-metallicity BT-Settl  models \citep{Baraffe+Chabrier+Barman+etal_2003, Allard+Homeier+Freytag_2011} to relate mass and age to $T_\mathrm{eff}$ and $\log g$. We interpolate between models using second-order splines in the logarithm of all the parameters, and we occasionally linearly extrapolate (in the logarithm of the parameters) the grid at low temperatures ($\lesssim300$~K) for masses $\lesssim6~M_\mathrm{Jup}$, with most of the extrapolation being necessary for masses in the range $2-4~M_\mathrm{Jup}$.

When characterizing substellar companions with direct-imaging observations, the system's age is one of the most important parameters. The reason for this is that, at fixed luminosity, age and mass are highly degenerate. In the limit of perfect knowledge of a companion's age, the prior $dp/dT_\mathrm{eff}/d\log g$ will be a line in the $T_\mathrm{eff}-\log g$ plane (an isochrone). In this case, uncertainties in evolutionary models will dominate the priors. As the age becomes increasingly uncertain, the prior is spread over an increasing range of temperatures and gravities, with the details of the distribution being determined by the physics of substellar cooling and the underlying mass distribution. 

The age of the substellar companion $\kappa$ And b is uncertain by nearly an order of magnitude, falling in the range ${\sim}30-200$~Myr \citep{Carson+Thalmann+Janson+etal_2013, Hinkley+Pueyo+Faherty+etal_2013, Bonnefoy+Currie+Marleau+etal_2014}. Even more uncertain, age estimates for GJ~504b range from as young as ${\sim}100$~Myr \citep{Kuzuhara+Tamura+Kudo+etal_2013} to as old as ${\sim}5$~Gyr \citep{Fuhrmann+Chini_2015}.  These age constraints lead to broad, but far from uniform, priors on $T_\mathrm{eff}$ and $\log g$. In Figure~\ref{fig:prior}, we show $dp/dT_\mathrm{eff}/d\log g$ for observations of a system with age constraints similar to  $\kappa$~And~b ($30-200$~Myr, middle panel) and GJ~504b ($100~\mathrm{Myr}-5$~Gyr, top panel). The mass ranges we assume for each calculation, which are indicated in the figure, also span likely ranges for these companions. The peak probability of each distribution is normalized to unity. In both panels, dark blue and white correspond to the peak and zero probability, respectively, but the color scales are otherwise different.

The ripple-like features that are present in the top and middle panels of Figure~\ref{fig:prior} are due to the onset of deuterium burning. Generically, low-temperature objects have the highest prior probability. We also see that the prior probability of $\log g$ at fixed $T_\mathrm{eff}$ generally increases with the value of $\log g$. Thus, if the effective temperature of a substellar companion is all that is known, it is more likely to have a high surface gravity. This trend is the result of the nonuniform cooling rates of substellar objects as a function of age (objects cool more rapidly at early times). As $\log g$ (at fixed temperature) increases, objects tend to fall on older isochrones. Since we assume $dp/d\log M\sim$ constant and the radii of these objects are only weak functions of mass and age, we have $dp / d\log g\sim$ constant. Therefore, the prior probability of $\log g$ at fixed temperature depends primarily on how fast objects cool past that temperature, resulting in prior probabilities that increase with $\log g$. 
 
The priors shown in the top and middle panels of Figure~\ref{fig:prior} both assume poor age constraints. In the bottom panel, we contrast this assumption with an example of a tight age constraint. This age range ($17-25$~Myr) is consistent with current estimates \citep{Binks+Jeffries_2014, Mamajek+Bell_2014} for the substellar companion $\beta$ Pic b \citep{Lagrange+Gratadour+Chauvin+etal_2009, Lagrange+Bonnefoy+Chauvin+etal_2010} and is more representative of the current sample of imaged companions including the HR8799 planets \citep{Marois+Macintosh+Barman+etal_2008, Marois+Zuckerman+Konopacky+etal_2010}, HD~95086~b \citep{Rameau+Chauvin+Lagrange+etal_2013}, 51~Eri~b \citep{Macintosh+Graham+Barman+etal_2015}, and HD~131399Ab \citep{Wagner+Apai+Kasper+etal_2016}. With such a well-constrained age, the details of the model become very important. Therefore, in addition to the BT-Settl ``hot-start'' model, we show priors for a series of  \citet[][SB12]{Spiegel+Burrows_2012} ``cold/warm-start'' models. We assume SB12's solar-metallicity ``hybrid cloud'' models; we also performed the calculations at $3\times$ solar metallicity and with their cloud-free models and found the priors to be dominated by the initial entropy. At such a young and precise age, uncertainties in the initial conditions dominate the priors. 

To demonstrate the effect of nonuniform priors on recovered $T_\mathrm{eff}$ and $\log g$ values, we perform retrievals on the ``true'' parameter values indicated as white stars in Figure~\ref{fig:prior}. These parameters are consistent with current estimates for GJ~504b, $\kappa$ And b, and $\beta$ Pic b. We include spectral covariance in each retrieval, and for the noise model, we use our LOCI noise parameters. Additionally, we assume $\mathrm{S/N}=10$ in the bottom and middle panels and $\mathrm{S/N}=4$ in the top panel---again, representative of plausible values for observations of these companions. The dashed black contours show the 95\% confidence region associated with each retrieval, where we have ignored the effect of the prior (i.e., using only the likelihood function). The solid red contours show the 95\% confidence region in a Bayesian parameter retrieval, which includes the effect of the prior (i.e., using the posterior probability distribution). The insets magnify where the contours and priors overlap, clarifying the probability variation within these regions. 

In the top panel of Figure~\ref{fig:prior}, where we have assumed an age constraint similar to estimates for GJ~504b, the prior probability varies by about an order of magnitude within the 95\% confidence contours. In this case, including the prior shifts the confidence region up in $\log g$ by ${\sim}$0.1~dex. In the middle panel, where we have assumed age constraints consistent with $\kappa$~And~b, the prior probability varies by a factor of ${\sim}5$ within the 95\% confidence contours. The dominant effect of including the prior in this case is to sharply cut off the confidence region at the edge of the prior's boundary. Uncertainties in evolutionary models dominate the priors shown in the bottom panel, where we have assumed a well-constrained age similar to $\beta$~Pic~b. In this case, our mock retrieval confirms the finding of previous studies that $\beta$~Pic~b's combination of age, surface gravity, and effective temperature is incompatible with cold-start initial conditions \citep{Lagrange+Bonnefoy+Chauvin+etal_2010, Quanz+Meter+Kenworthy+etal_2010}.

\section{Summary and Conclusions}\label{sec:conclusion}

In this paper, we have shown how to measure spectral errors and covariances in IFS data and demonstrated the importance of using the full covariance matrix, as opposed to assuming independent errors, in atmospheric parameter retrievals. By measuring the spectral errors and covariances in GPI early science data, we generate a realistic noise model with parameterizations for data PSF-subtracted with GPI's implementation of the LOCI and KLIP algorithms. We find that KLIP---whether its PSF approximation uses the full frame or 10 annuli---produces stronger large-scale spectral correlation than LOCI, which is likely due to subtle differences in the implementations of these algorithms (see Section~\ref{sec:parameterize}). This finding highlights the importance of measuring the spectral covariance for each data set and reduction technique.

We combine our noise model with theoretical exoplanet spectra \citep{Allard+Homeier+Freytag_2011} to perform mock retrievals of $T_\mathrm{eff}$ and $\log g$ with and without the full covariance matrix that we assume to be the source of the noise. In $5\times10^4$ realizations, the ``true'' parameter values in retrievals that ignore spectral covariance fell within the 95\% confidence region only ${\sim}33\%$ of the time for LOCI and ${\sim}23\%$ and of the time for KLIP. Scaling $\chi^2_\mathrm{dof}$ to unity worsens the problem, decreasing these fractions to ${\sim}26\%$ and ${\sim}14\%$, respectively. 

Finally, we explore the additional effect of nonuniform priors on recovered $T_\mathrm{eff}$ and $\log g$ values. Assuming flat priors on the age and logarithmic mass distributions of substellar companions, we generate prior probability distributions in the $T_\mathrm{eff}-\log g$ plane. Low-temperature companions have the highest prior probability, and the prior probability of a particular $\log g$ at fixed $T_\mathrm{eff}$ generally increases with the value of $\log g$. Thus, if the effective temperature of a substellar companion is all that is known, it is more likely to have a high surface gravity. We perform mock Bayesian parameter retrievals including spectral covariance on systems with and without well-constrained ages. For systems with well-constrained ages, we find that the prior probability can vary within the 95\% confidence region by as much as an order of magnitude. When the age is tightly constrained, the prior is dominated by uncertainties in evolutionary models. 

The new generation of high-contrast instruments are all IFSs, from P1640 \citep{Hinkley+Oppenheimer+Zimmerman+etal_2011}, GPI \citep{Ingraham+Marley+Macintosh+etal_2014}, SPHERE \citep{Beuzit+Feldt+Dohlen+etal_2008}, and CHARIS \citep{Peters-Limbach+Groff+Kasdin+etal_2013, Groff+Kasdin+Limbach+etal_2014} on large ground-based telescopes, to WFIRST-AFTA \citep{Spergel+Gehrels+Baltay+etal_2015} in space.  These new instruments will discover and obtain low-resolution spectra of faint exoplanets close to their host stars, where they would be undetectable without image processing.  While correlated noise is inevitable in such spectra, the methods we have presented enable it to be statistically characterized and properly included in Bayesian retrievals of atmospheric parameters.

\acknowledgments{The authors would like to thank the anonymous referee for a very helpful review that led to significant improvements in this paper. This research is based on observations obtained at the Gemini Observatory, which is operated by the Association of Universities for Research in Astronomy, Inc., under a cooperative agreement with the NSF on behalf of the Gemini partnership: the National Science Foundation (United States), the National Research Council (Canada), CONICYT (Chile), the Australian Research Council (Australia), Minist\'erio da Ci\^encia, Tecnologia e Inova\c c\~ao (Brazil) and Ministerio de Ciencia, Tecnolog\'ia e Innovaci\'on Productiva (Argentina). J.P.G. is supported by the National Science Foundation Graduate Research Fellowship under Grant No. DGE 1148900. This work was performed in part under contract with the Jet Propulsion Laboratory (JPL) funded by NASA through the Sagan Fellowship Program executed by the NASA Exoplanet Science Institute. J.P.G.~would like to thank Tim Morton for useful conversations.  The authors thank Ed Turner and Laurent Pueyo for helpful comments on the manuscript.}

\bibliographystyle{apj_eprint}
\bibliography{mybib}

\end{document}